\documentclass{article}
\usepackage{spconf,amsmath,amssymb,graphicx}
\usepackage{makecell}
\usepackage{microtype}
\usepackage{esint}
\usepackage{tikz}
\usetikzlibrary{positioning}
\usetikzlibrary{decorations.pathreplacing,calligraphy}
\ninept

\newcommand{\myfW}{\ensuremath{\boldsymbol{f}_{\mathbf{W}}}}
\newcommand{\myg}{\ensuremath{\boldsymbol{g}}}

\newcommand{\myM}{\ensuremath{\mathbf{M}}}
\newcommand{\mynabla}{\ensuremath{\boldsymbol{\nabla}}}

\newcommand{\myPhi}{\ensuremath{\boldsymbol{\Phi}}}
\newcommand{\myS}{\ensuremath{\mathbf{S}}}

\newcommand{\mytheta}{\ensuremath{\boldsymbol{\theta}}}
\newcommand{\myu}{\ensuremath{\boldsymbol{u}}}
\newcommand{\myW}{\ensuremath{\mathbf{W}}}
\newcommand{\myx}{\ensuremath{\boldsymbol{x}}}

\makeatletter
\newcommand{\bigcomp}{
  \DOTSB
  \mathop{\vphantom{\sum}\mathpalette\bigcomp@\relax}%
  \slimits@
}
\newcommand{\bigcomp@}[2]{%
  \begingroup\m@th
  \sbox\z@{$#1\sum$}%
  \setlength{\unitlength}{0.9\dimexpr\ht\z@+\dp\z@}%
  \vcenter{\hbox{%
    \begin{picture}(1,1)
    \bigcomp@linethickness{#1}
    \put(0.5,0.5){\circle{1}}
    \end{picture}%
  }}%
  \endgroup
}
\newcommand{\bigcomp@linethickness}[1]{%
  \linethickness{%
      \ifx#1\displaystyle 2\fontdimen8\textfont\else
      \ifx#1\textstyle 1.65\fontdimen8\textfont\else
      \ifx#1\scriptstyle 1.65\fontdimen8\scriptfont\else
      1.65\fontdimen8\scriptscriptfont\fi\fi\fi 3
  }%
}
\makeatother

\title{Perceptual--Neural--Physical Sound Matching}
\name{Han Han, Vincent Lostanlen, and Mathieu Lagrange}
\address{Nantes Université, École Centrale Nantes, CNRS, LS2N, UMR 6004, F-44000 Nantes, France}
\begin{document}
%
\maketitle
\begin{abstract}
Sound matching algorithms seek to approximate a target waveform by parametric audio synthesis.
Deep neural networks have achieved promising results in matching sustained harmonic tones.
However, the task is more challenging when targets are nonstationary and inharmonic, e.g., percussion.
We attribute this problem to the inadequacy of loss function. 
On one hand, mean square error in the parametric domain, known as ``P-loss'', is simple and fast but fails to accommodate the differing perceptual significance of each parameter.
On the other hand, mean square error in the spectrotemporal domain, known as ``spectral loss'', is perceptually motivated and serves in differentiable digital signal processing (DDSP).
Yet, spectral loss is a poor predictor of pitch intervals and its gradient may be computationally expensive; hence a slow convergence.
Against this conundrum, we present Perceptual-Neural-Physical loss (PNP).
PNP is the optimal quadratic approximation of spectral loss while being as fast as P-loss during training.
We instantiate PNP with physical modeling synthesis as decoder and joint time--frequency scattering transform (JTFS) as spectral representation.
We demonstrate its potential on matching synthetic drum sounds in comparison with other loss functions.
\end{abstract}
\section{Introduction}
\label{sec:intro}
Given an audio synthesizer \myg, the task of sound matching \cite{horner1995wavetable} consists in retrieving the parameter setting \mytheta{} that ``matches'' a target sound \myx; i.e., such that a human ear judges the generated sound $\myg(\mytheta)$ to resemble \myx.
Sound matching has applications in automatic music transcription, virtual reality, and audio engineering \cite{shier2021manifold,esling2019universal}. 
Of particular interest is the case where $\myg(\mytheta)$ solves a known partial differential equation (PDE) whose coefficients are contained in the vector \mytheta.
In this case, \mytheta{} reveals some key design choices in acoustical manufacturing, such as the shape and material properties of the resonator.

Over the past decade, the renewed interest for deep neural networks (DNN's) in audio content analysis has led researchers to formulate sound matching as a supervised learning problem \cite{gabrielli2018end}.
Intuitively, the goal is to optimize the synaptic weights $\mathbf{W}$ of a DNN \myfW{} so that $\myfW(\myx_n)=\tilde{\mytheta}_n$ approximates $\mytheta_n$ over a training set of pairs $(\myx_n, \mytheta_n)$.
Because $\myg$ automates the mapping from parameter $\mytheta_n$ to sound $\myx_n$, this training procedure incurs no real-world audio acquisition nor human annotation.
However, prior publications have pointed out that the approximation formula $\tilde{\mytheta}_n\approx\mytheta_n$ lacks a perceptual meaning:
depending on the choice of target $\myx_n$, some deviations $(\tilde{\mytheta}_n-\mytheta_n)$ may be judged to have a greater effect than others \cite{masuda2021synthesizer,roth2011a,yeeking2018}.

The paradigm of differentiable digital signal processing (DDSP) has brought a principled methodology to address this issue \cite{engel2020ddsp}.
The key idea behind DDSP is to chain the learnable encoder \myfW{} with the known decoder \myg{} and a non-learnable but differentiable feature map \myPhi{}.
In DDSP, \myfW{} is trained to minimize the perceptual distance between vectors $\myPhi(\tilde{\myx}_n) = (\myPhi \circ \myg \circ \myfW)(\myx_n)$ and $\myPhi(\myx_n)$ on average over samples $\myx_n$.
Yet, a practical shortcoming of DDSP is that it requires to backpropagate the ``spectral loss'' $\Vert \myPhi(\tilde{\myx}_n) - \myPhi(\myx_n)\Vert_2$ over each DNN prediction $\tilde{\mytheta}_n$; and so at every training step, since $\mathbf{W}$ is iteratively updated by stochastic gradient descent (SGD).

\begin{figure}
\begin{tikzpicture}[thick]
\node(theta){\mytheta};
\node(physical)[above=of theta, align=left, yshift=-9mm]{\scriptsize Parametric\\\scriptsize domain};
\node(x)[label={[anchor=west, xshift=-2mm, yshift=-5mm]\scriptsize  original}, right=of theta]{\myx};
\node(audio)[above=of x, align=left, yshift=-9mm]{\scriptsize Audio\\\scriptsize domain};
\node(S)[right=of x]{\myS};
\node(perceptual)[above=of S, align=left, yshift=-9mm]{\scriptsize Perceptual\\\scriptsize domain};

\node(thetatilde)[below=of theta]{$\tilde{\mytheta}$};
\node(xtilde)[label={[anchor=west, xshift=-2mm, yshift=-6mm]\scriptsize reconstruction}, right=of thetatilde]{$\tilde{\myx}$};
\node(Stilde)[below=of S]{$\tilde{\myS}$};

\draw[decorate, decoration = {calligraphic brace, mirror}] (3.1,-1.7) --  (3.1,0.25) node(DDSP)[xshift=7mm, pos=0.5, align=left] {\scriptsize  DDSP\\\scriptsize spectral$\;\approx$\\\scriptsize loss};

\draw[decorate, decoration = {calligraphic brace}] (5.5,-1.7) --  (5.5,0.25) node(PNP)[xshift=-6mm, pos=0.5, align=left]{\scriptsize PNP\\\scriptsize quadratic\\\scriptsize form};

\node(thetabis)[xshift=24mm, right=of S]{\mytheta};
\node(xbis)[right=of thetabis]{\myx};
\node(thetatildebis)[below=of thetabis]{$\tilde{\mytheta}$};

\node(M)[right=of DDSP,xshift=4mm]{$\myM(\mytheta)$};
\node(manifold)[above=of M, align=left, yshift=-2mm]{\scriptsize Riemannian\\\scriptsize metric};

\draw[->] (theta) -- (x) node[above,midway]{\myg};
\draw[->] (x) -- (S) node[above,midway]{\myPhi};
\draw[->, dashed] (x) -- (thetatilde) node[above,xshift=-2mm,midway]{\myfW};
\draw[->, dashed] (thetatilde) -- (xtilde) node[above,midway]{\myg};
\draw[->, dashed] (xtilde) -- (Stilde) node[above,midway]{\myPhi};

\draw[->] (thetabis) -- (xbis) node[above,midway]{\myg};
\draw[->, dashed] (xbis) -- (thetatildebis) node[above,xshift=-2mm,midway]{\myfW};
\draw[->] (thetabis) -- (M) node[above,xshift=-5mm,yshift=-1mm,midway]{$\mynabla_{(\myPhi\circ\myg)}$};


\end{tikzpicture}

\caption{Graphical outline of the proposed method.
Given a known synthesizer $\myg$ and feature map $\myPhi$, we train a neural network $\myfW$ to estimate $\tilde{\mytheta}$ and minimize the ``perceptual--neural--physical'' (PNP) quadratic form
$\langle\tilde{\mytheta} - \mytheta\big\vert\myM(\mytheta)\vert\tilde{\mytheta} - \mytheta\rangle$ where $\myM$ is the Riemannian metric associated to $(\myPhi \circ \myg)$.
Hence, PNP approximates DDSP spectral loss yet does not need to backpropagate $\mynabla_{(\myPhi\circ\myg)}(\tilde{\mytheta})$ at each epoch.
Transformations in solid (resp. dashed) lines can (resp. cannot) be cached during training.
\label{fig:overview}}
\end{figure}
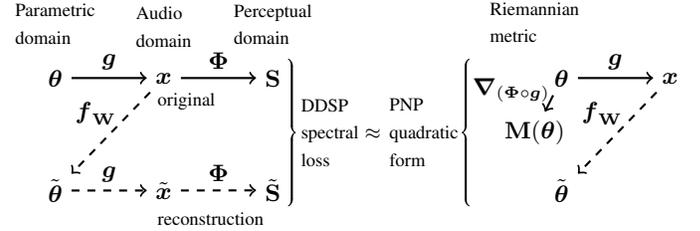

In this article, we propose a new learning objective for sound matching, named perceptual--neural--physical (PNP).
Our main contribution is to compute the Riemannian metric $\myM$ associated to the Jacobian $\mynabla_{(\myPhi\circ\myg)}$ over each sample $\mytheta_n$ (see Section \ref{sub:metric-learning}).
With $\myM(\mytheta_n)$, we train \myfW to minimize a locally linear approximation of spectral loss, making PNP comparable to DDSP.
Yet, unlike in DDSP, the computation of $\mynabla_{(\myPhi\circ\myg)}$ is independent from the encoder \myfW: thus, it may be parallelized and cached during DNN training.
A second novelty of our paper resides in its choice of application: namely, differentiable sound matching for percussion instruments.
This requires not only a fine characterization of the spectral envelope, as in the DDSP of sustained tones; but also of attack and release transients.
For this purpose, we need \myg{} and \myPhi{} to accommodate sharp spectrotemporal modulations.
Specifically, we rely on original differentiable implementations of the functional transformation method (FTM) for \myg{} and the joint time--frequency scattering transform (JTFS) for \myPhi. \footnote{Companion website: https://github.com/lylyhan/perceptual\_neural\_physical}
\footnote{Audio examples: https://pnp.cargo.site/}

\section{Methods}

\subsection{Accelerating spectral loss with Riemannian geometry}
\label{sub:metric-learning}

We assume the synthesizer $\myg$ and the feature map $\myPhi$ to be continuously differentiable.
Let us denote by $\mathcal{L}^{\mathrm{DDSP}}$ the ``spectral loss'' associated to the triplet $(\myPhi, \myfW, \myg)$.
Its value at a parameter set \mytheta{} is:
\begin{align}
    \mathcal{L}^{\mathrm{DDSP}}_{\mytheta}(\mathbf{W}) &=
    \dfrac{1}{2}\Vert\myPhi(\tilde{\myx}) - \myPhi(\myx)\Vert^2_2
    \nonumber \\
    &=
    \dfrac{1}{2}\big\Vert(\myPhi \circ \myg \circ \myfW \circ \myg)(\mytheta) - (\myPhi \circ \myg)(\mytheta)
    \big\Vert^2_2
\end{align}
by definition of $\tilde{\myx}$ and $\myx$.
Using $\tilde{\mytheta}$ as shorthand for $(\myfW\circ\myg)(\mytheta)$, we conduct a first-order Taylor expansion of $(\myPhi\circ\myg)$ near $\mytheta$. We obtain:
\begin{equation}
    \myPhi(\tilde{\myx}) =
    \myPhi(\myx) +
    \mynabla_{(\myPhi\circ\myg)}(\mytheta) \cdot
    (\tilde{\mytheta} - \mytheta) +
    O(\Vert \tilde{\mytheta} - \mytheta \Vert^2_2),
    \label{eq:phi-xtilde}
\end{equation}
where the Jacobian matrix $\mynabla_{(\myPhi\circ\myg)}(\mytheta)$ contains $P = \dim \myPhi(\myx)$ rows and $J = \dim \mytheta$ columns.
The manifold formed by differentiable map $(\myPhi\circ\myg)$ and the open set $\mytheta \subset \mathbb{R}^{J}$ induces a Riemannian metric $\myM$, i.e., an inner product on the tangent space at each point $\mytheta$:
\begin{equation}
    \myM(\mytheta)_{j,j^{\prime}} =
    \sum_{p=1}^{P}
    \left(\mynabla_{(\myPhi\circ\myg)}(\mytheta)_{p,j}\right)
    \left(\mynabla_{(\myPhi\circ\myg)}(\mytheta)_{p,j^{\prime}}\right).
    \label{eq:riemannian-metric}
\end{equation}
The real-valued square matrix $\myM(\mytheta)$ 
defines a positive semidefinite kernel which, once plugged into Equation \ref{eq:phi-xtilde}, serves to approximate $\mathcal{L}^{\mathrm{DDSP}}_{\mytheta}(\myW)$ in terms of a quadratic form over $(\tilde{\mytheta}-\mytheta)$:
\begin{equation}
    \Vert
    \myPhi(\tilde{\myx}) -
    \myPhi(\myx)
    \Vert_2^2 =
    \big\langle
    \tilde{\mytheta} - \mytheta
    \big\vert
    \myM(\mytheta)
    \big\vert
    \tilde{\mytheta} - \mytheta
    \big\rangle
    + O\big(\Vert \tilde{\mytheta} - \mytheta \Vert^3_2\big).
    \label{eq:quadratic-form}
\end{equation}

The advantage of the approximation above is that the metric \myM{} may be computed over the training set once and for all.
This is because Equation \ref{eq:riemannian-metric} is independent of the encoder \myfW.
Furthermore, since \mytheta{} is low-dimensional, we may store  $\myM(\mytheta)$ on RAM.
From this perspective, we define the perceptual--neural--physical  loss (PNP) associated to $(\myPhi,\myfW,\myg)$ as the linearization of spectral loss at \mytheta:
\begin{align}
    \mathcal{L}_{\mytheta}^{\mathrm{PNP}}(\myW) &= \dfrac{1}{2} 
    \big\langle(\myfW\circ\myg)(\mytheta) - \mytheta
    \big\vert\myM(\mytheta)\big\vert
    (\myfW\circ\myg)(\mytheta) - \mytheta\big\rangle
    \nonumber \\
    &=
    \mathcal{L}_{\mytheta}^{\mathrm{DDSP}}(\myW)
    +
    O\big(\Vert(\myfW\circ\myg)(\mytheta)-\mytheta\Vert^3_2\big).
    \label{eq:PNPloss}
\end{align}
According to the chain rule, the gradient of PNP loss at a given training pair $(\myx_n, \mytheta_n)$ with respect to some scalar weight $\myW_i$ is:
\begin{align}
    \dfrac{\partial \mathcal{L}^{\mathrm{PNP}}_{\mytheta}}{\partial \myW_i}(\mytheta_n)
    =
    \Big\langle
    \myfW(\myx_n) - \mytheta_n
    \Big\vert \myM(\mytheta_n) \Big\vert
    \dfrac{\partial \myfW}{\partial \myW_i}(\myx_n)
    \Big\rangle.
    \label{eq:grad-pnp}
\end{align}
Observe that replacing $\myM(\mytheta_n)$ by the identity matrix in the equation above would give the gradient of \emph{parameter loss} (P-loss); that is, the mean squared error between the predicted parameter $\tilde{\mytheta}$ and the true parameter $\mytheta$.
Hence, we may regard PNP as a perceptually motivated extension of P-loss, in which parameter deviations are locally recombined and rescaled so as to linearly approximate a DDSP objective.

The matrix $\mathbf{M}(\mytheta)$ is constant in $\mathbf{W}$.
Hence, its value may be cached across training epochs, and even across hyperparameter settings of the encoder.
In comparison with P-loss, the only computational overhead of PNP is the bilinear form in Equation \ref{eq:grad-pnp}.
However, this computation is performed in the parametric domain, i.e., in low dimension ($J = \dim \mytheta$).
Hence, its cost is negligible in front of the forward ($\myfW$) and backward pass ($\partial\myfW/\partial\myW_i$) of DNN training.

\subsection{Damped least squares}
\label{sub:LMA}

The principal components of the Jacobian $\mynabla_{(\myPhi\circ\myg)}(\mytheta)$ are the eigenvectors of $\mathbf{M}(\mytheta)$.
We denote them by $\boldsymbol{v}_j$ and the corresponding eigenvalues by $\sigma_j^2$: for each of them, we have $\myM(\mytheta) \boldsymbol{v}_j = \sigma^2_j \boldsymbol{v}_j$.
The $\boldsymbol{v}_j$'s form an orthonormal basis of $\mathbb{R}^J$, in which we can decompose the parameter deviation $(\tilde{\mytheta} - \mytheta)$.
Recalling Equation \ref{eq:PNPloss}, we obtain an alternative formula for PNP loss:
\begin{equation}
    \mathcal{L}_{\mytheta}^{\mathrm{PNP}}(\myW) = \dfrac{1}{2}
    \sum_{j=1}^{J} \sigma_j^2
    \left\vert
    \langle
    (\myfW\circ\myg)(\mytheta) - \mytheta
    \big\vert
    \boldsymbol{v}_j
    \rangle
    \label{eq:pnpevec}
    \right\vert^2
\end{equation}
The eigenvalues $\sigma_j^2$ stretch and compress the error vector along their associated direction $\boldsymbol{v}_j$, analogous to the magnification and suppression of perceptually relevant and irrelevant parameter deviations.
In practice however, when $\sigma_j^2$ cover drastic ranges or contain zeros, as presented below in Section \ref{sub:discussion}, the error vector is subject to extreme distortion and potential instability due to numerical precision errors. These scenarios, commonly referred to as $\myM$ being ill-conditioned, can lead to intractable learning objective $\mathcal{L}_{\theta}^{\mathrm{PNP}}$.

Reminiscent of the damping mechanism introduced in Levenberg-Marquardt algorithm when solving nonlinear optimization problems, we update Equation \ref{eq:PNPloss} as
\begin{align}
    \mathcal{L}_{\mytheta}^{\mathrm{PNP}}(\myW) &= \dfrac{1}{2} 
   \big\langle\tilde{\mytheta{}} - \mytheta
    \big\vert\myM(\mytheta)+\lambda I\big\vert
   \tilde{\mytheta} - \mytheta\big\rangle
    \label{eq:PNPloss_LMA}
\end{align}
The damping term $\lambda I$ up-shifts all eigenvalues of $\myM{}$ by a constant positive amount $\lambda$, thereby changing its condition number. 
At the limit of $\lambda \rightarrow 0$, $\mathcal{L}_{\theta}^{\mathrm{PNP}}$ reduces to a quadratic form which is asymptotically equivalent to spectral loss as P-loss approaches zero. At the limit of $\lambda\rightarrow \infty$, $\myM$ is negligible in front of $\lambda I$ thus $\mathcal{L}_{\theta}^{\mathrm{PNP}}$ boils down to P-loss.  
Alternatively, Equation \ref{eq:PNPloss_LMA} may also be viewed as a L2 regularization with coefficient $\lambda$.

To further address potential convergence issues, $\lambda$ may be scheduled or adaptively changed according to epoch validation loss. We adopt delayed gratification mechanism to decrease $\lambda$ by a factor of 5 when epoch validation loss is going down, and fix $\lambda$ otherwise.

\section{Application to drum sound matching}
\subsection{\emph{Perceptual: }Joint time--frequency scattering (JTFS)}
The joint time--frequency scattering transform (JTFS) is a nonlinear convolutional operator which extracts spectrotemporal modulations in the constant-$Q$ scalogram \cite{anden2019joint,andreux2020kymatio}.
Its kernels proceed from a separable product between two complex-valued wavelet filterbanks, defined over the time axis and over the log-frequency axis respectively.
After convolution, we apply pointwise complex modulus and temporal averaging to each JTFS coefficient.
These coefficients are known as scattering ``paths'' $p$.
We apply a logarithmic transformation to the feature vector $\mathrm{JTFS}(\myx_n)$ corresponding to each sound $\myx_n$, yielding
\begin{equation}
    \myS_{n,p} = \myPhi(\myx_n)_p = (\myPhi\circ\myg)(\mytheta_n)_p =
    \log\left(1+\dfrac{\mathrm{JTFS}(\myx_n)_p}{\varepsilon}\right),
\end{equation}
We set $\varepsilon=10^{-3}$, which is the order of magnitude of the median value of $\mathrm{JTFS}$ across all examples $\myx_n$ and paths $p$.

The multiresolution structure of JTFS is reminiscent of spectrotemporal receptive fields (STRF), and thus may serve as a biologically plausible predictor of neurophysiological responses in the primary auditory cortex \cite{chi2005multiresolution}.
At a higher level of music cognition, a recent study has shown that Euclidean distances in $\myPhi$ space predict auditory judgments of timbre similarity within a large vocabulary of instrumental playing techniques, as collected from a group of professional composers and non-expert music listeners \cite{lostanlen2021time}.

We use the GPU implementation of \cite{muradeli2022differentiable} 
to compute JTFS with the same parameters as \cite{lostanlen2021time}: $Q_1=12$, $Q_2=1$, and $Q^{\mathrm{fr}}=1$ filters per octave respectively. 
We set the temporal averaging to $T=3$ seconds and the frequential averaging to $F=2$ octaves; hence a total of $P=20762$ paths.
We refer to \cite{vahidi2023mesostructures} for further details on the ability of $\nabla_{(\myPhi\circ\myg)}$ to extract ``mesostructures'' in nonstationary audio signals.

\subsection{\emph{Neural: }Deep convolutional network (convnet)}

EfficientNet is a convolutional neural network architecture that balances the scaling of the depth, width and input resolution of consecutive convolutional blocks \cite{tan2019efficientnet}.
Achieving state-of-the-art performance on image classification with significantly less trainable parameters, its most light-weight version EfficientNet-B0 also succeeded in benchmarking audio classification tasks \cite{zeghidour2021leaf}. 
We adopt EfficientNet-B0 as our encoder $\myfW{}$, resulting in 4M learnable parameters. We append a linear dense layer of $J=\dim \mytheta{}$ neurons and a 1D batch normalization before tanh activation. The goal of batch normalization is to gaussianize the input, such that the activated output is capable of uniformly cover the normalized prediction range.
The input to $\myfW{}$ is the log-scaled CQT coefficients of each example, spanning 10 octaves with 12 filters per octave.

\subsection{\emph{Physical: }Functional transformation method (FTM)}
We are interested in the perpendicular displacement $\mathbf{X}(t,u)$ on a rectangular drum face, which can be solved from the following partial differential equation defined in the Cartesian coordinate system $\myu=(u_1,u_2)$.

\begin{align}   
    \left(\dfrac{\partial^2 \mathbf{X}}{\partial t^2}(t,\myu)
    - c^2 \nabla^2\mathbf{X}(t,\myu) \right)
    &+ S^4 \big(\nabla^4
    \mathbf{X}(t,\myu)\big)
    \nonumber \\
    +\dfrac{\partial}{\partial t}
    \Big(d_1 \mathbf{X}(t,\myu) + d_3 &\nabla^2\mathbf{X}(t,\myu)\Big) =
    0
    \label{eq:pde}
\end{align}

In addition to the standard traveling wave equation in the first above parenthesis, the fourth-order spatial and first-order time derivatives incorporate damping factors induced by stiffness, internal friction in the drum material and air friction in the external environment, rendering the solution a closer simulation to reality.
Specifically, $\alpha$, $S$, $c$, $d_1$, $d_3$ designate respectively the side length ratio, stiffness, traveling wave speed, frequency-independent damping and frequency-dependent damping of the drum. 
Even though real world drums are mostly circular, a rectangular drum model is equally capable of eliciting representative percussive sounds in real world scenarios. The circular drum model simply requires a conversion of Equation \ref{eq:pde} into the Polar coordinate system.
We bound the four sides of this $l$ by $l\alpha$ rectangular drum at zero at all time. 
For simplicity, we simulate the excitation by setting the initial condition at $t_0=0$ to be $\mathbf{X}(t_0,\myu=(0.4l, 0.4l\alpha)) = 0.03 \text{ meters}, \text{and } 0 \text{ otherwise}$.

We implement generator \myg{} as a PDE solver to this high-order damped wave equation, namely the functional transformation method (FTM) \cite{trautmannbookftm2003,schafer2019}. 
FTM solves the PDE by transforming the equation into its Laplace and functional space domain, where an algebraic solution can be obtained. It then finds the time-space domain solution via inverse functional transforms, expressed in an infinite modal summation form
\begin{align}
    \myx(t) = \mathbf{X}(t,\myu)
    = 
    \sum_{m\in\mathbb{N}^2}
    K_m(\myu,t) \exp(\sigma_m t) \sin(\omega_{m} t)
\end{align}
The coefficients $K_m(\myu,t)$, $\sigma_m$, $\omega_m$ are derived from the original PDE parameters in the following ways.
\begin{equation}
    \omega_m^2 = (S^4-\frac{d_3^2}{4})\Gamma_{m_1,m_2}^2 +
    (c^2+\frac{d_1d_3}{2})\Gamma_{m_1,m_2}
    - \frac{d_1^2}{4}
\end{equation}
\begin{equation}
     \sigma_m = \frac{d_3}{2}\Gamma_{m_1,m_2}-\frac{d_1}{2}
\end{equation}
\begin{equation}
     K_m(\myu,t) = y_{u}^{m}\delta(t)\sin(\frac{\pi m_1 u_1}{l})\sin\left(\frac{\pi m_2 u_2}{l\alpha}\right)
\end{equation}
where $\Gamma_{m_1,m_2}=\pi^2m_1^2/l^2+\pi^2 m_2^2/(l\alpha)^2 $, and $y_{u}^m$ is the $m^{th}$ coefficient associated to the eigenfunction $\sin(\pi m \myu/l)$ that decomposes $y_u(\myu)$.

Without losing connections to the acoustical manufacturing of the drum yet better relating \myg{}'s input with perceptual dimensions, we reparametrize the PDE parameters $\{S,c,d_1,d_3,\alpha\}$ into $\theta = \{\log\omega_1, \tau_1, \log p, \log D, \alpha\}$, detailed in Section 3.4 of \cite{han2020wav2shape}.
We prescribe sonically-plausible ranges for each parameter in $\theta$, normalize them between $-1$ and $1$, uniformly sample in the hyper-dimensional cube, and obtain a dataset of 100k percussive sounds sampled at 22050 HZ. The train/test/validation split is $8:1:1$.

In particular, fundamental frequency $\omega_1$, duration $\tau_1$ falls into ranges $[40,1000]$ Hz and $[0.4,3]$ seconds respectively. Inhomogeneous damping rate $p$, frequential dispersion $D$ and aspect ratio $\alpha$ ranges are $[10^{-5},0.2]$, $[10^{-5},0.3]$, and $[10^{-5},1]$.

\begin{figure}
    \centering
    \includegraphics[width=\linewidth]{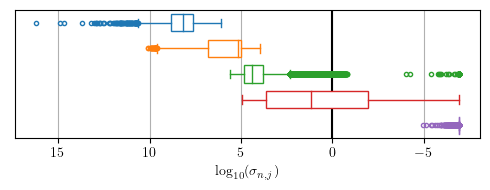}
    \caption{Distributions of the five sorted eigenvalues of $\mathbf{M}(\mytheta_n)$.
    For the sake of comparison between PNP and P-loss, the bold line indicates the eigenvalues of the identity matrix (see Equation \ref{eq:grad-pnp}).}
    \label{fig:eigenvalues}
\end{figure}

\section{Results}
\subsection{Baselines}
\label{subsec:baseline}
We train $f_W$ with 3 different losses - multi-scale spectral loss \cite{steinmetz2020auraloss}, parameter loss, and PNP loss. 
We use a batch size of 64 samples for spectral loss, and 256 samples for P-loss and PNP loss. The training proceeds for 70 epochs, where around $20\%$ of the training set is seen at each epoch. We use Adam optimizer with learning rate $10^{-3}$. Table \ref{tab:results} reports the training time per epoch on a single Tesla V100 16GB GPU.

\subsection{Evaluation with JTFS-based spectral loss}
We propose to use the L2 norm of JTFS coefficients error averaged over test set for evaluation. 
As a point of reference, we also include the average multi-scale spectral error, implemented as in Section \ref{subsec:baseline}.
One of the key distinctions between Euclidean JTFS distance and multi-scale spectral error is the former's inclusion of spectro-temporal modulations information.
Meanwhile unlike mean squared parameter error, both metrics reflect the perceptual closeness instead of parametric retrieval accuracy for each proposed model. 



\begin{table*}[ht]
    \centering
    \begin{tabular}{ll|lrrr}
    \makecell{Loss}
     &\makecell{\myPhi{}}
    &\makecell{Pitch} & \makecell{JTFS distance \\ (avg. on test set)}  &\makecell{MSS\\ (avg. on test set)} & \makecell{Training time \\ per epoch} \\ 
    \hline
    P-loss & --- & Known & $\boldsymbol{22.23}$ $\pm$ 2.17 & $\boldsymbol{0.31}$ $\pm$ 0.013 & 49 minutes \\
    $\mathcal{L}_{\theta}^{\mathrm{DDSP}}$ & $\myPhi_{\mathrm{MSS}}$ & Known & 31.86 $\pm$ 0.332  & 0.335 $\pm$ 0.005 & 54 minutes\\
    $\mathcal{L}_{\theta}^{\mathrm{PNP}}$ & $\myPhi_{\mathrm{JTFS}}$ & Known & 23.58 $\pm$ 0.877 & 0.335 $\pm$ 0.005 & 49 minutes \\
    $\mathcal{L}_{\theta}^{\mathrm{DDSP}}$ & $\myPhi_{\mathrm{JTFS}}$ &--- & --- & ---& \makecell{est., $>1$ day}\\
     P-loss & --- & Unknown & 61.91 $\pm$ 6.26 & 1.02 $\pm$ 0.094 & 53 minutes \\
    $\mathcal{L}_{\theta}^{\mathrm{DDSP}}$ & $\myPhi_{\mathrm{MSS}}$ & Unknown & 138.95 $\pm$ 37.12 & 1.59 $\pm$ 0.307 & 59 minutes\\
    $\mathcal{L}_{\theta}^{\mathrm{PNP}}$ & $\myPhi_{\mathrm{JTFS}}$ & Unknown & $\boldsymbol{61.21}\pm 1.207$ & $\boldsymbol{0.97} \pm 0.019$ & 49 minutes \\
    
    \end{tabular}
    \caption{Report of average JTFS distance and MSS metrics evaluated on test set. Six models are trained with two modalities: 1. the inclusion of pitch retrieval i.e. regressing  $\theta=\{\tau, \log p, \log D, \alpha\}$ vs. $\theta=\{\log \omega_1, \tau, \log p, \log D, \alpha\}$, and 2. the choice of loss function: P-loss, MSS loss, or PNP loss with adaptive damping mechanism.
    The best performing models with known and unknown pitch are P-loss and PNP loss respectively. Training with MSS loss is more time consuming than training with P-loss or PNP loss. Training with differentiable JTFS loss is unrealistic in the interest of time. }
    \label{tab:results}
\end{table*}

\subsection{Discussion}
\label{sub:discussion}
Despite being the optimal quadratic approximation of spectral loss, it is nontrivial to apply the bare PNP loss form as Equation \ref{eq:PNPloss} in experimental settings. On one hand, $\myPhi \circ \myg$ potentially has undesirable property that exposes the Riemannian metric calculations to numerical precision errors. On the other hand, extreme deformation of the optimization landscape may lead to the same numerical instability facing stochastic gradient descent with spectral loss.
We report on a few remedies that helped stabilize learning with PNP loss, and offer insights on future directions to take. 

First and foremost, our preliminary experiments show that training PNP loss without damping $\lambda=0$ subjects to convergence issues due to the high condition numbers in empirical $\myM$s as illustrated in Section \ref{sub:LMA}.
 Fig. \ref{fig:eigenvalues} shows the sorted eigenvalue distribution of all $\myM$s in test set, where $\myM$s are rank-2,3 or 4 matrices with eigenvalues ranging from $0$ to $10^{20}$. This could be an implication that entries of $\mytheta$ contain implicit linear dependencies in generator $\myg$, or that local variations of certain $\mytheta$ fail to linearize differences in the output of $\myg$ or $\myPhi \circ \myg$. As an example, the aspect ratio $\alpha$ influences the modal frequencies and decay rates via \cite[Equations 12--13]{han2020wav2shape}, where in fact $(1/\alpha + 1/\alpha^2)$ could be a better choice of variable that linearizes $\myg$.

To address $\myM$s' ill conditions we attempted at numerous damping mechanisms to update $\lambda$: namely, constant $\lambda$, scheduled $\lambda$ decay, and adaptive $\lambda$ decay. The intuition is to have $\mathcal{L}_{\theta}^{\mathrm{PNP}}$ start in the parameter loss regime and move towards the spectral loss regime while training. The best performing model is achieved with adaptive $\lambda$ decay (see Section \ref{sub:LMA}).
We propose to divide $\lambda$ by a factor of 5 if the model breaks the best epoch validation loss record and keep it the same otherwise. 
In practice, we initialize $\lambda$ to match the largest empirical $\sigma_j^2 \approx 10^{20}$, and then adaptively decay it to $3\times10^{14}$ in 20 epochs. This indicates that $\myfW$ is able to learn with damped PNP loss if $\lambda$ is large enough to compensate for rank deficiency in $\myM$. 
\begin{figure}
    \centering
    \includegraphics[width=\linewidth]{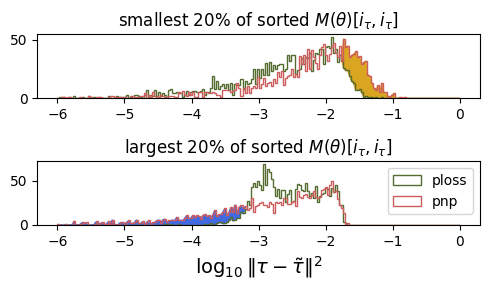}
    \caption{Histogram of log squared $\tau$ estimation error for perceptually significant (largest $20\%$ $\myM(\mytheta)[i_{\tau},i_{\tau}]$) and less significant (smallest $20\%$ $\myM(\mytheta)[i_{\tau},i_{\tau}]$) sounds. The yellow and blue regions indicate the weighting-induced difference in retrieval accuracy.}
    \label{fig:pnp-vs-loss}
\end{figure}

The diagonal elements of $\myM(\mytheta)$ can be regarded as both the applied weights' magnitudes and proxies for the perceptual importance of  $\theta$'s accuracy.
Inspecting the results of $\tau = \mytheta[i_{\tau}]$ regression, we observe in Fig. \ref{fig:pnp-vs-loss} that in comparison with P-loss model, PNP model improves retrieval accuracy for sounds inducing larger perceptual difference to changes in $\tau$ (in blue), at the expense of lowered accuracy for the opposite (in yellow). This suggests a trade-off behavior aligned with PNP loss' weighting scheme.

We believe that more of PNP loss' mathematical potential can be exploited in the future, notably in cases where parameterization without domain-specific knowledge renders the failure of P-loss, and its use in hybrid optimization schemes. 
We plan to investigate the scalability of each loss function under reparameterizations, as well as other damping schemes and optimizers. 
The current update mechanism, originated from the Leverberg-Marquardt algorithm, aims to improve the conditioning of a matrix inversion problem in the Gauss-Newton algorithm. However when used jointly with stochastic gradient descent, each $\lambda$ update may change the optimization landscape drastically. The resulting optimization behavior is thus not fully understood. 
We consider interfacing nonlinear least squares solver with SGD and forming a hybrid learning scheme in future work.




\section{Conclusion}
\label{sec:conclusion}
Knowledge on human auditory perception aiding data-driven approaches to machine listening tasks have been exemplified in a multitude of applications \cite{icassp23specialsession}. 
In this article we presented another case of this synergy named Perceptual-Neural-Physical (PNP) autoencoding, a bilinear form learning objective for sound matching task. 
In our application, PNP optimizes the retrieval of physical parameters from sounds in a perceptually-motivated metric space, enabled by differentiable implementations of physical model and computational proxy of neurophysiological construct of human auditory system.

We demonstrated PNP's mathematical relationship to spectral loss and parameter loss. Using this formulation, we motivated and established one way of interpolating between optimizing in parameter and spectral loss regimes. 
We presented damping mechanisms to facilitate its learning under ill-conditioned empirical settings and provided future plans for further exploiting its mathematical potential.

\clearpage 
\vfill\pagebreak

\bibliographystyle{IEEEbib}
\bibliography{han2023icassp}

\begin{thebibliography}{10}

\bibitem{horner1995wavetable}
Andrew Horner,
\newblock ``Wavetable matching synthesis of dynamic instruments with genetic
  algorithms,''
\newblock {\em Journal of the Audio Engineering Society}, vol. 43, no. 11, pp.
  916--931, 1995.

\bibitem{shier2021manifold}
Jordie Shier, Kirk McNally, George Tzanetakis, and Ky~Grace Brooks,
\newblock ``Manifold learning methods for visualization and browsing of drum
  machine samples,''
\newblock {\em Journal of the Audio Engineering Society}, vol. 69, no. 1/2, pp.
  40--53, 2021.

\bibitem{esling2019universal}
Philippe Esling, Naotake Masuda, Adrien Bardet, Romeo Despres, Axel Chemla,
  et~al.,
\newblock ``Universal audio synthesizer control with normalizing flows,''
\newblock in {\em Proceedings of the International Conference on Digital Audio
  Effects (DAFX)}, 2019.

\bibitem{gabrielli2018end}
Leonardo Gabrielli, Stefano Tomassetti, Carlo Zinato, and Francesco Piazza,
\newblock ``End-to-end learning for physics-based acoustic modeling,''
\newblock {\em IEEE Transactions on Emerging Topics in Computational
  Intelligence}, vol. 2, no. 2, pp. 160--170, 2018.

\bibitem{masuda2021synthesizer}
Naotake Masuda and Daisuke Saito,
\newblock ``Synthesizer sound matching with differentiable {DSP},''
\newblock in {\em Proceedings of the International Society on Music Information
  Retrieval (ISMIR) Conference}, 2021, pp. 428--434.

\bibitem{roth2011a}
Martin Roth and Matthew Yee-king,
\newblock ``A comparison of parametric optimization techniques for musical
  instrument tone matching,''
\newblock {\em Journal of the Audio Engineering Society}, May 2011.

\bibitem{yeeking2018}
Matthew Yee-King, Leon Fedden, and Mark d'Inverno,
\newblock ``Automatic programming of vst sound synthesizers using deep networks
  and other techniques,''
\newblock {\em IEEE Transactions on Emerging Topics in Computational
  Intelligence}, vol. 2, pp. 150--159, 2018.

\bibitem{engel2020ddsp}
Jesse Engel, Lamtharn~(Hanoi) Hantrakul, Chenjie Gu, and Adam Roberts,
\newblock ``{DDSP: Differentiable Digital Signal Processing},''
\newblock in {\em Proceedings of the International Conference on Learning
  Representations (ICLR)}, 2020.

\bibitem{anden2019joint}
Joakim And{\'e}n, Vincent Lostanlen, and St{\'e}phane Mallat,
\newblock ``Joint time--frequency scattering,''
\newblock {\em IEEE Transactions on Signal Processing}, vol. 67, no. 14, pp.
  3704--3718, 2019.

\bibitem{andreux2020kymatio}
Mathieu Andreux, Tom{\'a}s Angles, Georgios Exarchakis, Roberto Leonarduzzi,
  Gaspar Rochette, Louis Thiry, John Zarka, St{\'e}phane Mallat, Joakim
  And{\'e}n, Eugene Belilovsky, Joan Bruna, Vincent Lostanlen, Muawiz
  Chaudhary, Matthew~J. Hirn, Edouard Oyallon, Sixin Zhang, Carmine Cella, and
  Michael Eickenberg,
\newblock ``{Kymatio: Scattering transforms in Python},''
\newblock {\em Journal of Machine Learning Research}, vol. 21, no. 60, pp.
  1--6, 2020.

\bibitem{chi2005multiresolution}
Taishih Chi, Powen Ru, and Shihab~A Shamma,
\newblock ``Multiresolution spectrotemporal analysis of complex sounds,''
\newblock {\em The Journal of the Acoustical Society of America}, vol. 118, no.
  2, pp. 887--906, 2005.

\bibitem{lostanlen2021time}
Vincent Lostanlen, Christian El-Hajj, Mathias Rossignol, Gr{\'e}goire Lafay,
  Joakim And{\'e}n, and Mathieu Lagrange,
\newblock ``Time--frequency scattering accurately models auditory similarities
  between instrumental playing techniques,''
\newblock {\em EURASIP Journal on Audio, Speech, and Music Processing}, vol.
  2021, no. 1, pp. 1--21, 2021.

\bibitem{muradeli2022differentiable}
John Muradeli, Cyrus Vahidi, Changhong Wang, Han Han, Vincent Lostanlen,
  Mathieu Lagrange, and George Fazekas,
\newblock ``Differentiable time-frequency scattering in kymatio,''
\newblock in {\em Proceedings of the International Conference on Digital Audio
  Effects (DAFX)}, 2022.

\bibitem{vahidi2023mesostructures}
Cyrus Vahidi, Han Han, Changhong Wang, Mathieu Lagrange, Gy{\"o}rgy Fazekas,
  and Vincent Lostanlen,
\newblock ``Mesostructures: {B}eyond spectrogram loss in differentiable
  time-frequency analysis,''
\newblock {\em arXiv preprint arXiv:2301.10183}, 2023.

\bibitem{tan2019efficientnet}
Mingxing Tan and Quoc Le,
\newblock ``{EfficientNet: Rethinking model scaling for convolutional neural
  networks},''
\newblock in {\em Proceedings of the International conference on Machine
  Learning (ICML)}. PMLR, 2019, pp. 6105--6114.

\bibitem{zeghidour2021leaf}
Neil Zeghidour, Olivier Teboul, F{\'e}lix de~Chaumont~Quitry, and Marco
  Tagliasacchi,
\newblock ``{LEAF}: {A} learnable frontend for audio classification,''
\newblock {\em ICLR}, 2021.

\bibitem{trautmannbookftm2003}
L.~Trautmann and Rudolf Rabenstein,
\newblock {\em Digital Sound Synthesis by Physical Modeling Using the
  Functional Transformation Method},
\newblock Springer, 2003.

\bibitem{schafer2019}
M.~Schäfer, M.~Werner, and R.~Rabenstein,
\newblock ``Physical modeling in sound synthesis: Vibrating plates,''
\newblock in {\em Proc. 26th International Congress on Sound and Vibration
  (ICSV26)}, Montreal, Canada, Jul. 2019, pp. 1--8.

\bibitem{han2020wav2shape}
Han Han and Vincent Lostanlen,
\newblock ``{wav2shape: Hearing the Shape of a Drum Machine},''
\newblock in {\em Proceedings of Forum Acusticum}, 2020, pp. 647--654.

\bibitem{steinmetz2020auraloss}
Christian~J. Steinmetz and Joshua~D. Reiss,
\newblock ``auraloss: {A}udio focused loss functions in {PyTorch},''
\newblock in {\em Digital Music Research Network One-day Workshop (DMRN+15)},
  2020.

\bibitem{icassp23specialsession}
Laurie~M. Heller, Benjamin Elizalde, Bhiksha Raj, and Soham Deshmukh,
\newblock ``{Synergy between human and machine approaches to sound/scene
  recognition and processing: An overview of ICASSP special session},''
\newblock in {\em Proceedings of the IEEE International Conference on Audio,
  Speech, and Signal Processing}. 2023, IEEE.

\end{thebibliography}

\end{document}